\def\bm#1{\mbox{\boldmath $#1$}}
\documentclass[aps,pra,twocolumn,groupedaddress]{revtex4}
\usepackage{graphicx}

\begin{document}


\title{Laser-induced splittings in the nuclear magnetic resonance spectra of the rare gases}


\author{Rodolfo H. Romero}
\email{rodolfo@chem.helsinki.fi}
\altaffiliation{Permanent address: Department of Physics, Facultad de Ciencias Exactas y Naturales,
Universidad Nacional del Noredeste, Avenida Libertad 5500, (3400)
Corrientes, Argentina. {\em email:} {\tt rhromero@exa.unne.edu.ar}}
\affiliation{Laboratory for Instruction in Swedish, Department of Chemistry, P.O.~Box~55, A.~I.~Virtasen aukio 1, FIN-00014 University of Helsinki, Finland}
\author{Juha Vaara}
\email{jvaara@chem.helsinki.fi}
\affiliation{Laboratory of Physical Chemistry, Department of Chemistry, P.O.~Box~55, A.~I.~Virtasen aukio 1, FIN-00014 University of Helsinki, Finland}


\date{\today}

\begin{abstract}
Circularly polarized laser field causes a shift in the nuclear
magnetic resonance (NMR) spectra of all substances. The shift is
proportional to the intensity of the laser beam and yields oppositely
signed values for left- and right-circularly polarized light, CPL -/+,
respectively.  Rapid switching --- in the NMR time scale --- between
CPL+ and CPL- gives rise to a splitting of the NMR resonance lines.
We present uncorrelated and correlated quadratic response calculations
of the splitting per unit of beam intensity in the NMR spectra of
$^{21}$Ne, $^{83}$Kr, and $^{129}$Xe. We study both the regions far
away from and near to optical resonance and predict off-resonance
shifts of the order 0.01, 0.1, and $1\times 10^{-6}$~Hz for $^{21}$Ne,
$^{83}$Kr, and $^{129}$Xe, respectively, for a beam intensity of
10~W/cm$^2$. Enhancement by several orders of magnitude is predicted
as the beam frequency approaches resonance. Only then can the effect
on guest $^{129}$Xe atoms be potentially useful as a probe of the
properties of the host material.
\end{abstract}


\maketitle

Nuclear magnetic resonance (NMR) spectroscopy has become one of the
most successful techniques for the analysis of molecular
structure~\cite{Slichter}. Irradiation by circularly polarized light
(CPL) from a laser
has been proposed as a potentially useful technique of enhancing the
resolution of NMR spectra~\cite{Evans91}.  Early experiments searching
for the effect, in a solution of chiral molecules, reported shifts of
the order of 1~Hz in a 270~MHz NMR spectrum~\cite{Warren92}.  Much of
that magnitude can, however, be explained by heating effects and only
a residual of at most 0.1~Hz could be interpreted as originating in
other mechanisms~\cite{Warren98}.  Theory for the laser field-induced
NMR shifts was presented by Buckingham and Parlett~\cite{Buckingham94,
Buckingham97}, in terms of the inverse Faraday effect~\cite{van der
Ziel65}, {\em i.e.}, induced magnetization caused by the CPL beam
propagating through any sample. Since the laser-induced shifts become
proportional to the square of the electric field of the laser or,
equivalently, the intensity of the beam~\cite{Harris94, Buckingham94,
Buckingham97}, initial proposals involved using high-intensity pulsed
lasers to obtain shifts of the order of GHz~\cite{Evans91}. This is,
however, ruled out in NMR experiments, even leaving aside the heating
problem, due to the long lifetime of the nuclear spin states.  Hence,
intensities of the order of tens of W/cm$^2$, obtainable from
continuous wave lasers, must be considered in estimates of the order
of magnitude of the induced shifts.  Left- and right-circularly
polarized light cause opposite shifts. If the laser field is switched
between the two modes rapidly in the NMR time scale, the spectral lines are
doubled, with splitting proportional to the beam intensity.

NMR of guest rare gas atoms can be used as a sensitive probe of the
microstructure of the surrounding medium~\cite{jok99}. In this work,
we present {\em ab initio\/} electronic structure calculations of the
laser-induced splitting per unit of beam intensity in the NMR spectra
of free atomic $^{21}$Ne, $^{83}$Kr, and $^{129}$Xe, both at
frequencies away from and near to optical resonance.  In the
off-resonance region, at standard laser wave lengths and intensities,
the splittings in $^{129}$Xe are below the limit of current
experimental capabilities, while a large enhancement is expected when
approaching resonance.

A CPL beam of frequency $\omega$ induces a magnetic field at the
position of the nucleus $K$ in a molecule given by~\cite{Buckingham94,
Buckingham97}
\begin{equation}
B^{\pm}_{K,\alpha} = \frac{1}{2\omega} b^{K}_{\alpha\beta\gamma}
          \left(E^\pm_\beta \dot{E}^\pm_\gamma - E^\pm_\gamma
          \dot{E}^\pm_\beta\right).
\end{equation}
Here, $\bm{E}^\pm$ represents the electric field of the right ($+$) or
left ($-$) circularly polarized beam, and $b^{K}_{\alpha\beta\gamma}$
can be calculated as a third-order perturbation expression
from~\cite{Jaszunski99}
\begin{equation}
b^{K}_{\alpha\beta\gamma}  
 = -\frac{\alpha^2}{2}\,{\rm Im}
   \langle\langle \sum_i\frac{\ell_{iK,\alpha}}{r_{iK}^3}; r_\beta, r_\gamma
   \rangle\rangle_{\omega,-\omega},
\label{quad resp}
\end{equation}
where we have used atomic units 
and the notation for the quadratic response function~\cite{Yarkony}
$\langle\langle A; B, C\rangle\rangle_{\omega_1,\omega_2}$, equivalent
to time-dependent third-order perturbation theory with the
time-dependent operators $B$ and $C$, as well as their respective
frequencies $\omega_1$ and $\omega_2$.  In Eq.~(\ref{quad resp}),
$\alpha$ is the fine structure constant, $r_\beta$ and $r_\gamma$ are
the components of the electric dipole moment, $r_{iK}$ is
the distance between the electron $i$ and the nucleus $K$, and
$\ell_{iK,\alpha}$ is the $\alpha$ component of the angular momentum
of the electron $i$ around the nucleus $K$.  While $-\alpha^2\sum_i
\bm{\ell}_{iK}/r_{iK}^3$ is the magnetic field at nucleus $K$ due to the
orbital motion of the electrons, Eq.~(\ref{quad resp}) corresponds to
the modification of this quantity by the external, time-dependent
electric field.

The relevant quantity for NMR experiments in the isotropic gas or
liquid phase is the isotropic rotational average
\begin{equation}
b_K = \frac{1}{6}\,\sum_{\alpha\beta\gamma}
\epsilon_{\alpha\beta\gamma}b^{K}_{\alpha\beta\gamma} 
= \frac{1}{3}\,\left(b_{xyz}^K + b_{yzx}^K + b_{zxy}^K\right),
\label{b_K isotropic}
\end{equation}
where $\epsilon_{\alpha\beta\gamma}$ stands for the Levi-Civita tensor
and $(x,y,z)$ is the molecule-fixed Cartesian frame. For spherically
symmetric systems, Eq.~(\ref{b_K isotropic}) reduces to one of the
components of $b^{K}_{\alpha\beta\gamma}$, {\em e.g.} $b^{K}_{xyz}$.
The induced field is stationary and oriented along the direction of
propagation of the beam. It couples to the magnetic moment $\gamma_K
\bm{I}_K$ of nucleus $K$, to give a term in the NMR spin Hamiltonian 
(in frequency units) in the high-field approximation as
\begin{equation}
H_{\rm NMR} = \pm\frac{1}{4\pi}\,\gamma_K I_{K,Z} b_{K} E_0^2,
\label{H_NMR}
\end{equation}
where $\gamma_K$ is the magnetogyric ratio of $K$, $I_{K,Z}$ is the
component of $\bm{I}_K$ along the external magnetic field, and $E_0$
is the amplitude of the electric field associated with the laser beam.
Eq.~(\ref{H_NMR}) corresponds to equally large but oppositely directed
frequency shifts, $\Delta$ and $-\Delta$, for the two differently
polarized beams.  
$\Delta$ depends on the intensity $I_0$ of the beam, and
becomes~\cite{units}
\begin{equation}
\Delta = \frac{1}{4\pi c\epsilon_0}\,\gamma_K b_{K} I_0,
\end{equation}
where $c$ and $\epsilon_0$ are the speed of light and permittivity of
vacuum, respectively.

Calculations of $\Delta/I_0$ were performed with the {\sc Dalton}
program~\cite{Dalton}, using the implementations of analytic quadratic
response functions of Ref.~\onlinecite{het92} at the {\em ab initio\/}
self-consistent field (SCF) and multiconfiguration SCF (MCSCF) levels,
and Ref.~\onlinecite{sal02} using density-functional theory (DFT).
We studied the basis set convergence of the laser-induced shifts,
starting with the uncontracted Gaussian basis set denoted by HIVu in
Ref.~\cite{Vaara03b}. Diffuse (small-exponent) functions were added to
each block of angular momentum until convergence of the results at the
correlated MCSCF level [using the complete active space
(CAS)~\cite{CASSCF} wave function specified in Table~\ref{wf}] to
within 0.1\% was reached.  This procedure has been shown to lead to
results close to the basis set limit in the calculation of nuclear
magnetic shielding constants of the rare gases~\cite{Vaara03a}.
The final, converged basis sets are $(15s\,11p\,7d\,5f)$,
$(19s\,16p\,15d\,5f)$, and $(26s\,20p\,16d\,3f)$ for Ne, Kr and Xe,
respectively. All the results reported below correspond to these
converged basis.

Both CAS and restricted active space (RAS)~\cite{RASSCF} -type MCSCF
wave functions were used.  The details of the chosen active spaces are
given in Table~\ref{wf}. The three-parameter hybrid B3LYP
functional~\cite{B3LYP} was used in the DFT calculations.  
The calculations were performed
for a set of typical laser frequencies. We also calculated the lowest
singlet excitation energies for each atom with a CASSCF wave function,
to estimate the range of frequencies where optical resonances occur.
These ranges are 0.63--1.05~a.u.\ for Ne, 0.41--0.52~a.u.\ for Kr, and
0.35--0.45~a.u.\ for Xe.  $\Delta/I_0$ was evaluated at frequencies
approaching the resonance, too.

The results for $\Delta/I_0$ in $^{21}$Ne, $^{83}$Kr and $^{129}$Xe
are presented in Table~\ref{shifts}. The range of the shifts,
calculated at SCF level, due to a laser beam intensity of 10~W/cm$^2$
and wavelengths between 13190 and 4880~\AA, is 1.7--4.7~nHz for
$^{21}$Ne, 9.6--30~nHz for $^{83}$Kr, and 80--290~nHz for
$^{129}$Xe. The corresponding values for $^{131}$Xe are of the
opposite sign and smaller than those for $^{129}$Xe, and can be
obtained from the latter by multiplying by the quotient of the
respective magnetogyric ratios $\gamma_{^{131}{\rm
Xe}}/\gamma_{^{129}{\rm Xe}}=-0.296$.

Inclusion of electron correlation at the CAS level increases
$\Delta/I_0$ by about about 15\% for Ne, but decreases it by {\em
ca.\/} 20\% for Kr and Xe. Comparison of the results obtained with the
different RAS and CAS wave functions shows that the choice of the
active space is more important than using the multireference (CAS) wave
functions.  With RAS wave functions, correlation increases
the shifts by up to around 15\% for Ne and 25\% for Kr and Xe,
compared to the SCF level. For all atoms and frequencies, the
DFT/B3LYP results are significantly larger than those corresponding to
the {\em ab initio\/} methods. We have also tested the LDA and BLYP
functionals, which resulted in even larger shifts.

The larger shifts obtained for Xe as compared to Ne and Kr can be
attributed to the larger polarizability of xenon. The electron cloud
is deformed in response to the external electric field, thus giving a
corresponding larger induced magnetic field at the nuclear site.
Results of correlated CAS calculations of $\Delta/I_0$ and the dynamic
polarizability $\alpha\!\left(\omega\right)$, for Ne, Kr, and Xe as a
function of the laser frequency $\omega$, are depicted in
Fig.~\ref{fig}. The frequency ranges from 0.1~a.u.\ to near the
threshold of optical resonance, where the properties diverge because
of the poles in linear and quadratic response functions.  The growth
of $\Delta/I_0$ with $\omega$, although qualitatively similar to
$\alpha\!\left(\omega\right)$, is faster than in the latter
property. There is an enhancement of $\Delta/I_0$ by several orders of
magnitude when optical resonance is approached.

Finally, we have also carried out a preliminary investigation of the
influence of relativistic effects on $\Delta$, by including the
mass-velocity $H^{\rm mv}=-\frac{1}{8}\,\alpha^2\sum_i
\nabla_i^4$, and Darwin $H^{\rm Dar}=\frac{1}{2}\,\alpha^2\pi\sum_K
Z_K\sum_i\delta\!\left(\bm{r}_{iK}\right)$
Hamiltonians as additional perturbations. The relativistic corrections
\begin{eqnarray}
b_K^{\rm mv} &=& -\frac{\alpha^2}{2}\,{\rm Im} \langle\langle \sum_i
             \frac{\ell_{iK,x}}{r_{iK}^3}; y, z, H^{\rm mv}
             \rangle\rangle_{\omega,-\omega,0}, \\
b_K^{\rm Dar} &=& -\frac{\alpha^2}{2}\,{\rm Im} \langle\langle \sum_i
             \frac{\ell_{iK,x}}{r_{iK}^3}; y, z, H^{\rm Dar}
             \rangle\rangle_{\omega,-\omega,0}.
\end{eqnarray}
were evaluated using cubic response functions~\cite{jon96} at the CAS
level.
The corresponding corrections to the shifts, $\Delta^{\rm mv}$ and
$\Delta^{\rm Dar}$, have opposite signs and partially cancel each
other. As expected, the relativistic effects in $^{21}$Ne are completely
negligible.
For $^{83}$Kr, $b_K^{\rm mv}$ and $b_K^{\rm Dar}$
represent, individually, corrections of the same order of magnitude as
the uncorrected value of $b_K$, although their partial cancellation 
finally leads to values approximately
25\% larger than the nonrelativistic value.
Finally, $b_{\rm Xe}^{\rm mv}$ and $b_{\rm Xe}^{\rm Dar}$ are larger
than the nonrelativistic values, roughly by a factor of three. Their
cancellation leads to a corrected value of about
$-50$~nHz for all the frequencies studied,
with $I_0 = 10$~W/cm$^2$.
No definitive conclusions may be drawn based on these results, however,
as the response functions retain their nonrelativistic pole structure
in this approach.  Furthermore, picture change effects~\cite{bar97} on
the hyperfine operator are also presently neglected.  A fully
relativistic quadratic response calculation would be more appropriate,
and will be pursued in the future.

In summary, we have calculated the shifts induced by circularly
polarized laser beam, to the NMR spectra for atomic $^{21}$Ne,
$^{83}$Kr, and $^{129}$Xe using first principles electronic structure
methods.
At typical beam intensities and laser frequencies, the shifts are much
too small for observation, with
the shift for $^{129}$Xe predicted at the order of magnitude of
1~$\mu$Hz. 
Experimental techniques exploiting higher beam frequencies would
benefit from the dramatic enhancement expected at near-resonant
frequencies. If realised experimentally, the effect may provide a new
characteristic signature of molecular structure in NMR spectroscopy.

M.~Jaszu\'nski (Warsaw), A.~Rizzo (Pisa), J.~Lounila (Oulu), and
J.~Jokisaari (Oulu) are thanked for useful discussions, as well as
T.~Helgaker (Oslo) and P.~Sa{\l}ek (Stockholm) for providing a
pre-release DFT version of the {\sc Dalton} software.  Financial
support from the Magnus Ehrnrooth Foundation (RHR), the Emil Aaltonen
Foundation (JV), and the Academy of Finland (RHR and JV, benefiting
also from the project 206001 of J.~Jokisaari) has been received. RHR
is on leave from the Universidad Nacional del Nordeste
(Argentina). Computational resources were partially provided by the
Center for Scientific Computing, Espoo, Finland.

			   
\clearpage
\begin{table}
\caption{\label{wf} Active atomic orbital spaces of the MCSCF wave functions used.}
\vspace{0.5cm}
\begin{ruledtabular}
\begin{tabular}{clcl}
 Atom    & Wave Function &  $N_e$$^a$  &  Active Space              \\  \hline
  Ne     &  CAS          &    8    & $2s2p\rightarrow 3s3p3d  $ \\
         &  RAS-I        &    8    & $2s2p\rightarrow 3s3p3d  $ \\
		 &  RAS-III      &   10    & $1s2s2p\rightarrow 3s3p3d4s4p4d4f5s5p  $ \\  \hline
  Kr     &  CAS          &    8    & $4s4p\rightarrow 4d$       \\
         &  RAS-I        &    8    & $4s4p\rightarrow 4d$       \\
		 &  RAS-II       &   18    & $3d4s4p\rightarrow 4d4f5s5p5d$ \\
		 &  RAS-III      &   26    & $3s3p3d4s4p\rightarrow 4d4f5s5p5d5f6s6p$ \\ \hline
  Xe     &  CAS          &    8    & $5s5p\rightarrow 5d$ \\
         &  RAS-I        &    8    & $5s5p\rightarrow 5d$  \\
		 &  RAS-II       &   18    & $4d5s5p\rightarrow 4f5d6s6p6d$ \\
		 &  RAS-III      &   26    & $4s4p4d5s5p\rightarrow 4f5d5f6s6p6d7s7p$ \\
\end{tabular}
\end{ruledtabular}
\begin{flushleft}
$^a$~$N_e$ is the number of correlated electrons.
\end{flushleft}
\end{table}
\begin{table}[H]
\caption{\label{shifts} Calculated laser-induced NMR line shifts per 
unit of laser beam intensity, $\Delta/I_0$ [in $10^{-9}$~Hz/(W$\,$cm$^{-2}$)]
in rare gas atoms Ne, Kr, and Xe.}
\vspace{0.5cm}
\begin{ruledtabular}
\begin{tabular}{llllllll}
 Nucleus    &$\omega$$^a$ (a.u.)
			                &   SCF   &  CAS  & RAS-I  & RAS-II & RAS-III  &  B3LYP  \\
\hline 
 $^{21}$Ne 	&   0.0345439   &  0.17   & 0.19   & 0.19   &   --   &  0.19    &  0.31   \\
			&   0.0428227   &  0.21   & 0.24   & 0.23   &   --   &  0.24    &  0.38   \\
			&   0.0656249   &  0.32   & 0.37   & 0.36   &   --   &  0.37    &  0.60   \\
			&   0.0773571   &  0.38   & 0.44   & 0.43   &   --   &  0.44    &  0.71   \\
			&   0.0885585   &  0.44   & 0.51   & 0.50   &   --   &  0.51    &  0.83   \\
			&   0.0932147   &  0.47   & 0.53   & 0.52   &   --   &  0.53    &  0.87   \\
\hline
  $^{83}$Kr &    0.0345439  &  0.96   & 0.80   &  0.80  &  0.85  &  1.49    &    1.76  \\
			&    0.0428227  &  1.17   & 1.01   &  1.01  &  1.06  &  1.86    &    2.24  \\
			&    0.0656249  &  1.92   & 1.60   &  1.60  &  1.76  &  2.98    &    3.62  \\
			&    0.0773571  &  2.34   & 1.97   &  1.97  &  2.13  &  3.57    &    4.42  \\
			&    0.0885585  &  2.82   & 2.34   &  2.34  &  2.56  &  4.26    &    5.27  \\
			&    0.0932147  &  2.98   & 2.50   &  2.50  &  2.77  &  4.53    &    5.70  \\
\hline
$^{129}$Xe	&	 0.0345439  &  7.99   & 5.32   &  5.32  &   9.58 &  10.65   &    19.17 \\
			&    0.0428227  & 10.11   & 6.92   &  6.92  &  12.25 &  13.31   &    24.50 \\
			&    0.0656249  & 17.04   & 12.25  &  12.25 &  20.24 &  22.90   &    41.00 \\
			&    0.0773571  & 21.83   & 15.44  &  15.44 &  25.56 &  28.22   &    51.12 \\
			&    0.0885585  & 26.62   & 19.17  &  19.70 &  31.42 &  34.61   &    62.84 \\
			&    0.0932147  & 29.29   & 20.77  &  21.30 &  34.08 &  37.81   &    68.16 \\
\end{tabular}
\end{ruledtabular}
\begin{flushleft}
$^a$~The frequencies correspond to wavelengths $\lambda=13190$, 10640,
			6943, 5890, 5145, and 4888 \AA, in the
			respective order.
\end{flushleft}
\end{table}


\clearpage
\noindent
{\bf FIGURE CAPTIONS}

\begin{figure}[h!]
\caption{\label{fig} Results of correlated CAS calculations of (a) the 
laser-induced NMR shift per unit of beam intensity $\Delta/I_0$, and (b)
frequency-dependent polarizability $\alpha\!\left(\omega\right)$, as a
function of the frequency $\omega$, for atomic Ne, Kr, and Xe.}
\end{figure}

\clearpage
\includegraphics[scale=0.75]{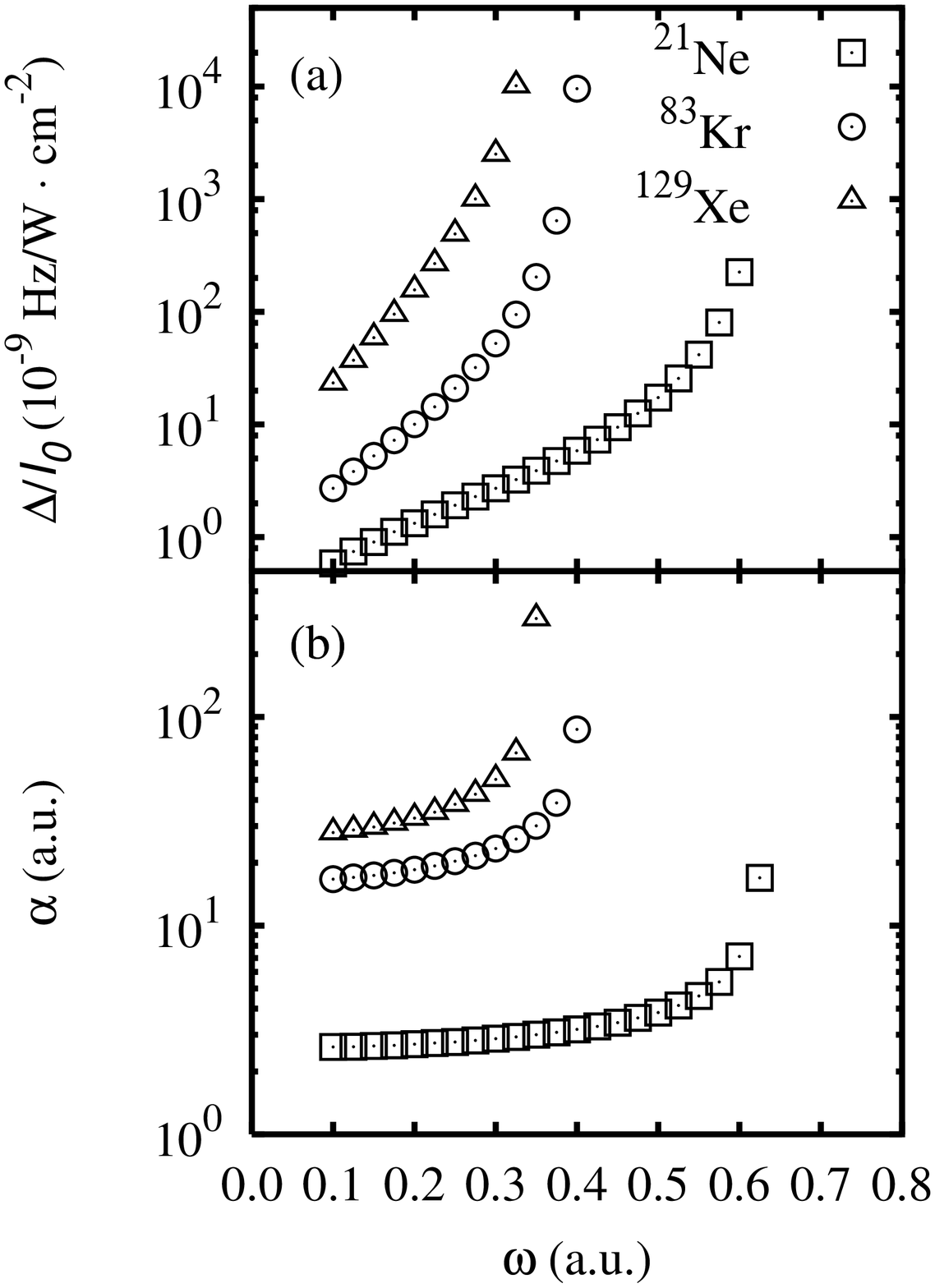}\\

\noindent
Fig.~\ref{fig}, Romero and Vaara, Phys.\ Rev.\ A.


\begin{thebibliography}{99}
\bibitem{units} To change from atomic
units to SI units we use the equivalence 1~a.u.\ of $b = m e
a_0^3/\alpha\hbar^2 c = 8.889238\times
10^{-19}$~s$\,$V$^{-1}$.
\end{thebibliography}
\end{document}